\documentclass[aps,prl,twocolumn,groupedaddress]{revtex4}
\usepackage{graphicx}
\usepackage{amsmath}

\begin{document}

\title{Two-mode squeezing in an electromechanical resonator}

\author{I. Mahboob}

\author{H. Okamoto}

\author{K. Onomitsu}

\author{H. Yamaguchi}


\affiliation{NTT Basic Research Laboratories, NTT Corporation, Atsugi-shi, Kanagawa 243-0198, Japan}



\maketitle

\textbf{The widespread availability of quantum entanglement with photons, in the guise of two-mode squeezed states, can be attributed to the phenomenon of parametric down-conversion \cite{knight, sqz2, sqz3, sqz4, sqz5}. A reinterpretation of this effect with macroscopic mechanical objects can offer a route towards a {\it purely mechanical} entanglement \cite{gs, sqz10} and the unique possibility of probing the quantum mechanical nature of our everyday classical world \cite{sqz11, sqz17}. In spite of this prospect, mechanical two-mode squeezed states have remained elusive due to the inability to recreate the nonlinear interaction at the heart of this phenomenon in the mechanical domain. To address this we have developed a parametric down-converter, in a mechanical resonator integrated with electrical functionality, which enables mechanical nonlinearities to be dynamically engineered to emulate the parametric down-conversion interaction. In this configuration, phonons are simultaneously generated in pairs in two macroscopic vibration modes which results in the amplification of their motion \cite{sqz07, cavmika}. In parallel, mechanical two-mode squeezed states are also created which exhibit fluctuations far below the thermal level of their constituent modes as well as harbouring correlations between the modes that become almost perfect as their amplification is increased \cite{sqz6, sqz7, sqz8, sqz9, sqz10}. This remarkable observation of correlations between two massive phonon ensembles paves the way towards an entangled macroscopic mechanical system at the single phonon level.}

Two-mode squeezed states pioneered in quantum optics offer a highly versatile resource for non-classical light, where entangled photon pairs are generated from parametric down-conversion in a non-linear media \cite{knight, sqz2, sqz3, sqz4, sqz5}. Indeed the success of two-mode squeezing has even inspired a reinterpretation of this concept with microwave photons by exploiting the Josephson non-linearity in superconducting circuits \cite{sqz07, sqz6, sqz7, sqz8, sqz9}. Perhaps the most striking iteration of this phenomenon can be found in hybrid systems composed of a mechanically compliant element that is parametrically coupled to an electromagnetic resonator \cite{cavopto}. These so-called cavity electro/opto-mechanical systems can readily host parametric down-conversion which not only amplifies both resonators \cite{cavmika} but at the single phonon/photon level it can generate an entanglement between two vastly dissimilar systems \cite{sqz10}. Tantalisingly, an all-mechanical variant of this interaction would not only open up a path towards the generation of a macroscopic mechanical entanglement \cite{sqz17, gs} but it would also offer the unique prospect of decoding the nature of its absence in our everyday classical world \cite{sqz11}.  

One intuitive approach to mechanical two-mode squeezing is to utilise multimode electromechanical resonators which can mimic the dynamics of cavity electro/opto-mechanical systems \cite{cpl3, imranK, co3, co4}. Although these phonon-cavity electromechanical systems can host parametric down-conversion, in practice, the large input excitation needed to activate the non-linear parametric interaction between two vibration modes has yielded only modest gains which has rendered the study of this interaction inaccessible \cite{cpl3, imranK}. To address this, we have developed an electromechanical resonator, shown in Fig. 1a and detailed elsewhere, that consists of two doubly clamped beams that are strongly inter-coupled via an exaggerated overhang between them \cite{sqz14}. This structure sustains two spectrally closely-spaced vibration modes labelled symmetric (S) and asymmetric (A), also shown in Fig. 1a, and in combination with piezoelectric transducers that are incorporated directly into the mechanical elements, provides the key to realising efficient parametric down-conversion.

\begin{figure*}[!hbt]
\begin{center}
\vspace{-0.5cm}
\includegraphics[scale=0.9]{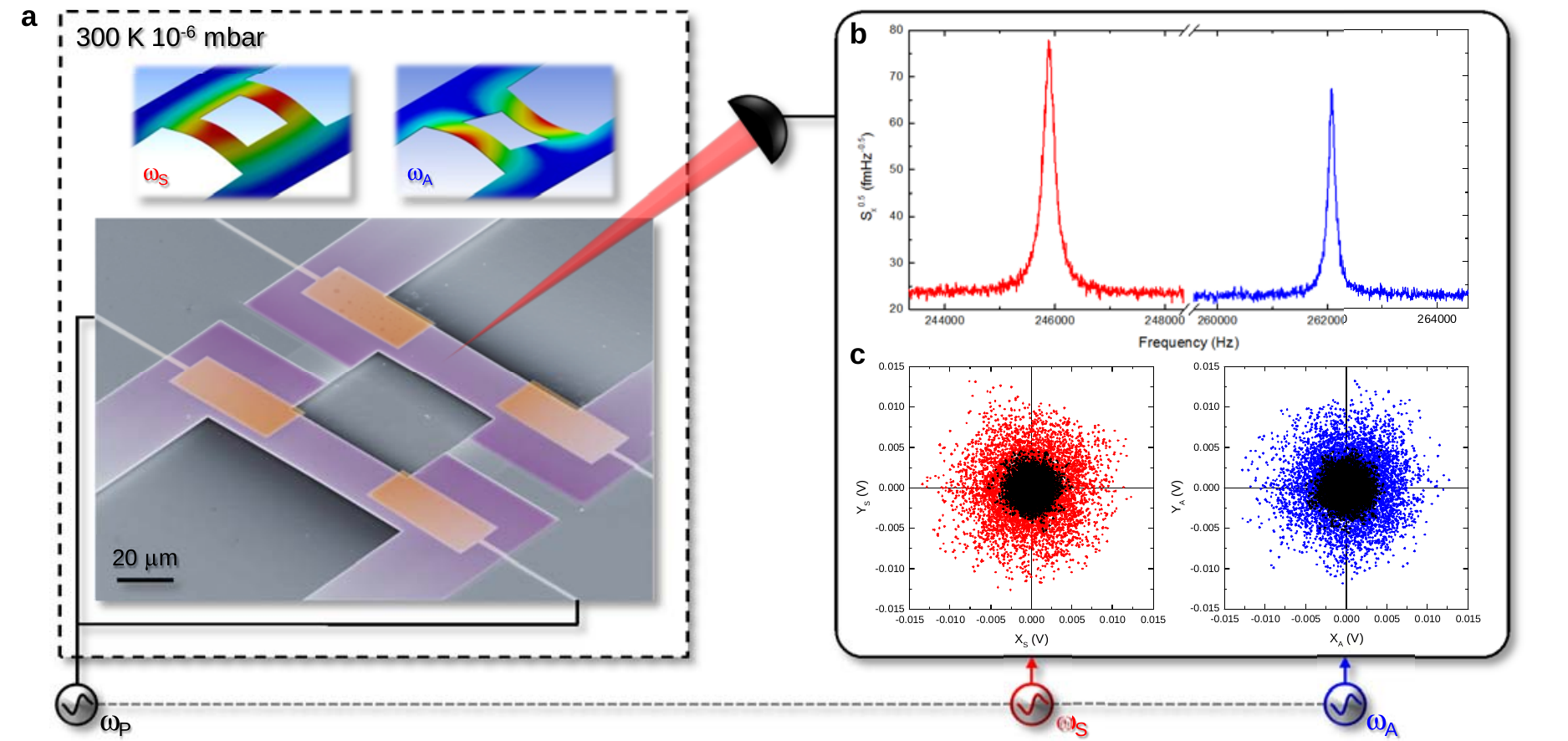}
\vspace{0.0cm}
\small{\caption{{\bf The non-degenerate electromechanical parametric amplifier.} {\bf a,} An electron micrograph of the GaAs based coupled electromechanical resonators incorporating a buried Si-doped layer that is confined within a shallow mesa (purple) where the mechanical elements are integrated with Au Schottky gates (orange) to form the piezoelectric transducers (see SI). The vibration of the modes is detected via the partially depicted laser interferometer from the right beam with the parametric down-conversion being piezoelectrically activated by applying a pump voltage to both electrodes on the left beam (see SI). The pumping generator ($\omega_{_P}$) and the local oscillators ($\omega_{_S}$ and $\omega_{_A}$) are all continuously synchronised (grey dashed-line) to ensure phase locked measurements. Also shown are the vibration profiles for the symmetric and asymmetric modes extracted from a finite element analysis. {\bf b,} The output noise from the interferometer, measured in a spectrum analyser, reveals the two modes via their picometre order thermal vibrations. {\bf c,} The on-resonance thermal motion of both modes projected in phase-space. Also shown is the electrical noise in the measurement setup (black points) acquired in an off-resonance configuration (see SI).}}
\end{center}
\end{figure*}

The piezoelectric transducers offer potent means to non-linearly modulate the spring constant of the electromechanical system by generating stress \cite{imranK, co3}. If this modulation is activated at the frequency difference between the modes, it enables a beam splitter interaction to be switched on, which results in sidebands being generated around both modes and their subsequent overlap enables them to couple and exchange energy \cite{imranK, co3, cav11, cav12}. On the other hand, if this modulation is implemented at the sum frequency of the two modes, it can permit parametric down-conversion of the input modulation, resulting in non-degenerate parametric amplification of both modes \cite{imranK, cavmika}. Physically, the sum frequency modulation of the electromechanical system's spring constant has components that can parametrically activate the symmetric and asymmetric modes. 

The total Hamiltonian of the system in this configuration can then be expressed as 

\vspace{-0.0cm}
\begin{eqnarray}
H = \sum_{n=S}^A \big(P_n^2/2m_n + m_n \omega_n^2 Q_n^2/2 \big) + \Lambda Q_{_S} Q_{_A}\mbox{cos}(\omega_{_P}t)
\end{eqnarray}

\noindent where the first two terms are the kinetic and potential energies respectively with canonical coordinates $Q_n$ and $P_n$ denoting the position and conjugate momentum of the constituent modes with frequencies $\omega_n$ and mass $m_n$ \cite{imranK}. The last term describes the parametric down-conversion when the modulation is activated at the sum frequency $\omega_{_P} = \omega_{_S} + \omega_{_A}$ where its amplitude $\Lambda$ is proportional to the pump voltage \cite{co3}. Although the present formulation is manifestly classically, the last term is analogous to $a_{_S}a_{_A}+a_{_S}^{\dagger}a_{_A}^{\dagger}$ in a quantum mechanical picture within the rotating frame approximation, where $a_n$ and $a_n^{\dagger}$ are the annihilation and creation operators for the two modes, and this interaction forms the basis of two-mode squeezing i.e. entanglement generation from parametric down-conversion \cite{sqz15, knight}. Ostensibly, entanglement between the two modes is unavailable in the present classically bound system, but remarkably this Hamiltonian indicates that correlations between the modes can emerge even in the limit of large phonon populations (see Supplementary Information (SI)). Consequently an observation of the correlations generated by the parametric down-conversion process, between two classical modes, would lay a pivotal marker on the road towards entangling two massive mechanical systems \cite{sqz17}.

The vibration of the two modes can be readily resolved via their thermal motion, in the output noise spectrum from an optical interferometer based probe, which reveals $\omega_{_S}/2\pi = 246$ kHz and $\omega_{_A}/2\pi = 262$ kHz with quality factors of 1300 and 2200 respectively as shown in Fig. 1b. Alternatively, the thermal motion from both modes can also be projected into phase-space by decomposing their displacements $Q_n(t)=X_n\mbox{cos}(\omega_nt) + Y_n\mbox{sin}(\omega_n t)$ within a narrow bandwidth into in-phase and quadrature components i.e. $X$ and $Y$ respectively. In practice this involves mixing the output from the interferometer with two local oscillators, which are locked exactly onto the resonances of the two modes, and then demodulated in two lock-in detectors over a period of 300 seconds with a sampling rate of 50 ms resulting in four simultaneously acquired time-series for all four components, each with 6000 points per measurement. The phase portraits reconstructed from this data yield circularly symmetric distributions, shown in Fig. 1c, which confirm that the thermal motion of both modes is random with all vibration phases being equally available \cite{rugars}.

\begin{figure*}[!hbt]
\begin{center}
\vspace{-0.5cm}
\includegraphics[scale=1.0]{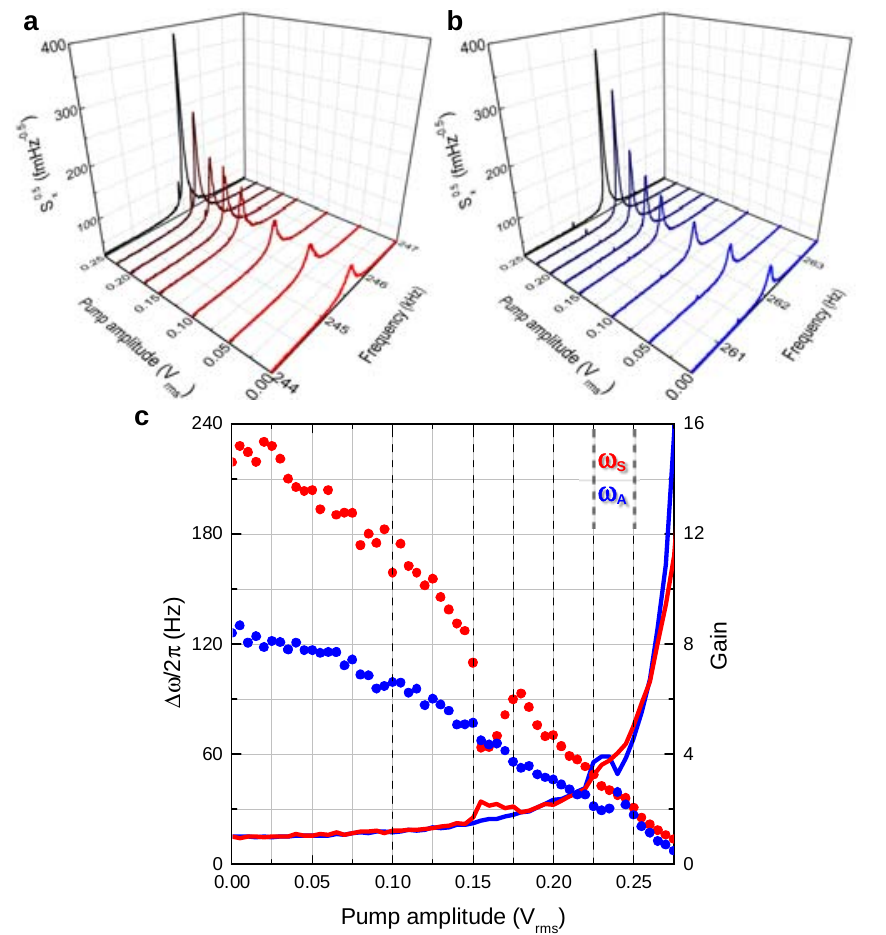}
\vspace{-0.0cm}
\small{\caption{{\bf Electromechanical non-degenerate parametric amplification.} {\bf a} and {\bf b,} The thermal motion of the symmetric and asymmetric modes respectively is amplified when the electromechanical resonator's spring constant is pumped at the sum frequency of the constituent modes. {\bf c,} The resultant gain of both modes referenced to their bare thermal motion (solid line) is accompanied with a narrowing of their respective power bandwidths $\Delta \omega/2\pi$ (points) as a function of the pump intensity. At the largest pump amplitudes, the quality factor of both modes converges to $\sim 10^5$ before undergoing non-degenerate parametric resonance. }}
\end{center}
\end{figure*}

First in order to investigate the availability of parametric amplification, namely the rate at which phonons are generated begins to exceed their rate of decay from both modes, the system is pumped at $\omega_{_P}$ by activating the piezoelectric transducers as detailed in Fig. 1a and the output noise spectra around both modes is acquired as shown in Figs. 2a and 2b. This measurement reveals that the thermal motion of both modes can be amplified as the pump voltage is increased, with gains of more than 20 dB becoming available, before they undergo regenerative oscillations as shown in Fig. 2c \cite{cavopto, cavmika, imranK, co3}. Repeating this measurement and projecting the outputs in phase-space, as detailed above and shown in Figs. 3a and 3b, not only reconfirms the amplification but it also indicates the phase preserving nature of this effect \cite{sqz07, sqz6}.

\begin{figure*}[!hbt]
\begin{center}
\vspace{-0.5cm}
\includegraphics[scale=0.9]{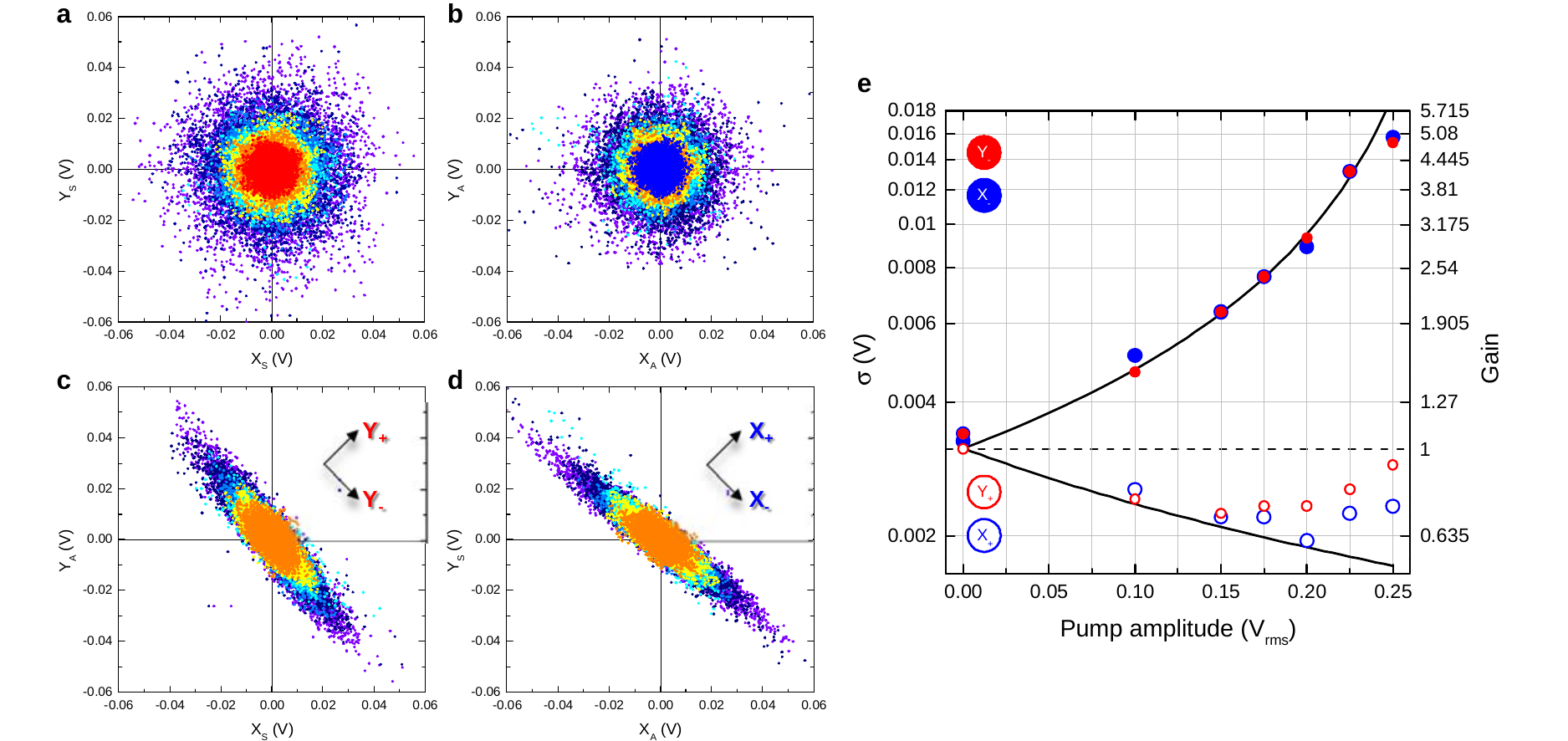}
\vspace{0.0cm}
\small{\caption{{\bf Two-mode squeezed states in an electromechanical resonator.} {\bf a} and {\bf b,} The phase portraits of the symmetric and asymmetric modes exhibit phase-conserving non-degenerate parametric amplification in response to the pump phonons undergoing parametric down-conversion when their intensity is increased from 0-0.25 V$_{rms}$ in increments detailed in Fig. 3e. {\bf c} and {\bf d,} The cross-quadratures of the above data take only specific values in phase-space indicating the existence of correlations between the modes where the {\it squeezing} is enhanced as the pump intensity is increased. {\bf e,} The squeezed states are quantified, in the rotated axis depicted in the insets to Figs. 3c and 3d, via the standard deviations of their phase-space distributions (points) where the gain is calibrated with respect to the narrowest bare thermal distribution from $Y_{_A}$ (dashed line) and is consistent with the theoretical trends derived from the above Hamiltonian (solid line) as detailed in SI. Note that the uncertainties associated with the extracted standard deviations are two orders of magnitude smaller and cannot be shown (see SI).}}
\end{center}
\end{figure*}

The parametric down-conversion simultaneously generates signal and idler phonons in both modes thus amplifying their thermal motion. As a result the noise, namely the amplified thermal motion of both modes, should be strongly correlated or in other words a two-mode squeezed state will be generated \cite{sqz6}. In order to confirm this, the cross-quadratures of the data shown in Figs. 3a and 3b are extracted i.e. the in-phase component of the symmetric mode versus the quadrature component of the asymmetric mode and vice versa are plotted as shown in Figs 3c and 3d. This analysis unambiguously reveals two-mode squeezing, where the noise along one phase is de-amplified at the expense of increased noise in the perpendicular phase, resulting in {\it squeezed} distributions. This specificity in phase-space indicates that the two modes are correlated whereas an absence of any correlations would plainly lead to circular distributions. 

In order to quantify the noise in the two-mode squeezed states, a phase factor is introduced to rotate and then project the amplified/squeezed distributions onto the rotated in-phase/quadrature axis, namely $X_{_-}$ and $Y_{_-}/X_{_+}$ and $Y_{_+}$ as shown in Figs. 3c and 3d, to statistically weigh their profiles (see SI). This analysis reveals Gaussian distributions with zero mean for all components where the extracted standard deviations $\sigma$ as a function of the pump amplitude enable a measure of the variation of the thermal fluctuations generated in the two-mode squeezed states as shown in Fig. 3e (see SI). As expected, the amplification of the in-phase components results in larger standard deviations with the corresponding gains consistent with the noise spectra measurements detailed by the dashed lines in Fig. 2c. Concurrently, the quadrature component distributions become narrower with their standard deviations becoming even less than that of the bare modes as referenced by the dashed line in Fig. 3e. Calibrating this reduction with respect to the narrowest bare thermal distribution yields a conservative lower limit of -4.76 dB squeezing below the thermal level at 300 K \cite{rugars, suh, sqz12} which is limited by pump induced heating and is detailed in SI.

\begin{figure*}[!hbt]
\begin{center}
\vspace{-0.5cm}
\includegraphics[scale=1.0]{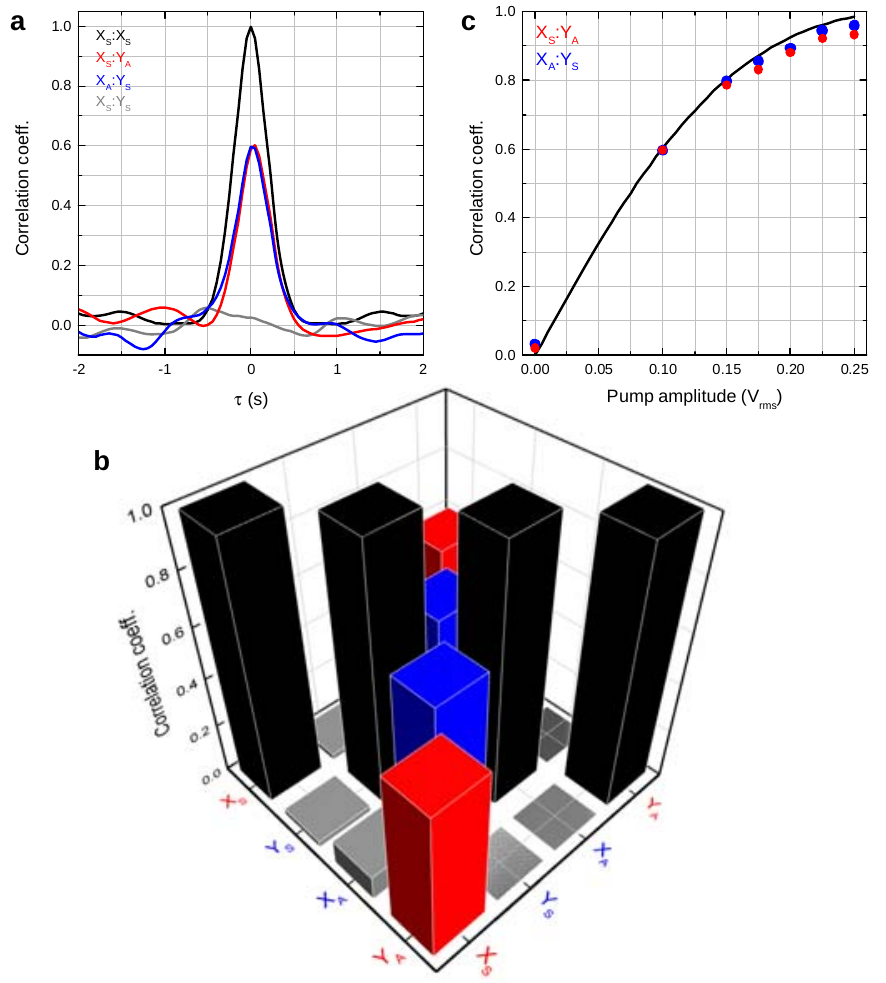}
\vspace{-0.0cm}
\small{\caption{{\bf Correlated macroscopic mechanical vibration modes.} {\bf a,} The absolute correlation coefficient between various combinations of components from both modes extracted with a pump amplitude of 0.1 V$_{rms}$ as a function of delay time $\tau$. {\bf b,} The corresponding correlation density matrix at $\tau$=0 reveals finite off-diagonal elements that verify the intertwined nature of the two-mode squeezed states. {\bf c,} The variation of the off-diagonal elements of the density matrix as a function of pump intensity (points) and the corresponding theoretical response (solid line). }}
\end{center}
\end{figure*}

Although noise reduction in the two-mode squeezed states below the thermal level of the bare modes is one characteristic of their interdependence \cite{sqz7}; further substantiating evidence can also be elicited by analysing the temporal correlations in the data shown in Figs. 3c-3d. To that end the absolute correlation coefficient $|COV(Z_iZ_j(\tau))/\sigma_{Z_i}\sigma_{Z_j}(\tau)|$, where the numerator describes the covariance, $Z_i \in \{X_{_S}, Y_{_S}, X_{_A}, Y_{_A}\}$ and $\tau$ is a delay between the constituent time series, is evaluated. The results of this analysis for a pump amplitude of 0.1 V$_{rms}$, shown in Fig. 4a, reveal the absence of any correlation between $X_{_S}$ and $Y_{_S}$, as attested to by their circular distribution in phase-space (see Fig. 3a). In contrast, the auto-correlation of $X_{_S}$ is exactly 1 at $\tau = 0$, confirming the perfect correlation expected in this configuration. However, finite correlations between $X_{_S}:Y_{_A}$ and $X_{_A}:Y_{_S}$ can be also discerned at $\tau = 0$ which indicates that the pump simultaneously generates signal and idler phonons in both modes in pairs which is the signature feature of parametric down-conversion \cite{sqz16}. 

The correlations at $\tau$=0 can be recast into a density matrix for all permutations of $Z_i$, as shown in Fig. 4b, which reveals in addition to the expected perfect auto-correlations captured by the diagonal elements, finite valued off-diagonal elements which confirm the existence of correlations between the two modes. Naturally the magnitude of the off-diagonal elements can be amplified as the pump intensity is increased as shown in Fig. 4c and this quantitatively confirms that the two mechanical vibration modes become almost perfectly entwined as their correlation coefficient tends to 1. The corresponding variation of the correlation coefficient can be reproduced by the parametric down-conversion Hamiltonian as shown in Fig. 4c and thus the simultaneous generation of phonons in both modes not only amplifies their thermal fluctuations but in the process it fundamentally links them (see SI).

The observation of correlations between two ensembles of millions of phonons, corresponding to the tangible motion of two massive mechanical vibration modes, opens up a new perspective on the study of correlated mechanical systems as well as heralding the prospect of quantum-optics being translated to acoustics in mechanically compliant architectures \cite{rob, sqz18}. Consequently these results hail a landmark step on the journey towards generating an all-mechanical macroscopic entanglement as they establish the non-linear interaction at the heart two-mode squeezing can be accessed in a purely mechanical system.


\vspace{0.5cm}

\small{

\noindent \textbf{Acknowledgements} The authors are grateful to D. Hatanaka and N. Kitajima for support, N. Lambert, J. R. Johansson and F. Nori for discussions and comments. This work was partly supported by JSPS KAKENHI Grant Number 23241046.

\vspace{0.5cm}

\noindent \textbf{Author contributions} I.M. conceived the idea, performed the measurements, analysed the data, developed the numerical model and wrote the paper. H.O. designed and fabricated the electromechanical resonator. K.O. fabricated the GaAs based heterostructure. H.Y. developed the analytical model and planned the project.  

\vspace{0.5cm}

\noindent \textbf{Author Information} Correspondence and requests for materials should be addressed to I.M. (imran.mahboob@lab.ntt.co.jp).}




\end{document}